\PassOptionsToPackage{table,xcdraw,svgnames,dvipsnames}{xcolor}
\documentclass[9pt,sigconf,letterpaper]{acmart}

\usepackage{balance}

\newcommand{\ie}{i.e., \@}
\newcommand{\eg}{e.g., \@}

\newcommand{\eat}[1]{}

\newcommand{\curlie}[1]{Curlie}

\newcommand{\eid}{engine ID\xspace}
\newcommand{\Eid}{\expandafter\MakeUppercase \eid}

\newcommand{\Eids}{{\expandafter\MakeUppercase \eid}s\xspace}
\newcommand{\etime}{engine time\xspace}
\newcommand{\Etime}{\expandafter\MakeUppercase \etime}
\newcommand{\eboots}{engine boots\xspace}
\newcommand{\Eboots}{\expandafter\MakeUppercase \eboots}
\newcommand{\lastreboot}{last reboot time\xspace}
\newcommand{\Lastreboot}{\expandafter\MakeUppercase \lastreboot}

\usepackage[normalem]{ulem}
\usepackage{savesym}
\savesymbol{Bbbk}
\usepackage{tabularx}
\newcolumntype{L}[1]{>{\raggedright\arraybackslash}p{#1}}
\newcolumntype{C}[1]{>{\centering\arraybackslash}p{#1}}
\newcolumntype{R}[1]{>{\raggedleft\arraybackslash}p{#1}}
\usepackage{graphicx}
\usepackage{subfig}
\usepackage{float}
\usepackage{subfig}
\usepackage{enumitem}
\setlist{nolistsep}
\usepackage{graphicx}
\usepackage{epstopdf}
\usepackage{grffile}
\usepackage{amssymb}
\usepackage{amsmath}
\usepackage{rotating}
\usepackage{url}
\usepackage{enumitem}
\usepackage{color}
\usepackage{xspace}
\usepackage{ifpdf}
\usepackage{mdwlist}
\usepackage{colortbl}
\usepackage{caption}
\usepackage{amssymb}
\usepackage{booktabs}
\usepackage{multicol}
\usepackage{multirow}
\usepackage{bigdelim}
\usepackage{enumitem}
\usepackage{xfrac}
\usepackage{afterpage}\usepackage{pdflscape}
\usepackage{longtable}
\usepackage{hhline}
\usepackage{lscape}
\usepackage{pifont}
\usepackage{fancyvrb}

\usepackage{arydshln} 

\usepackage[absolute,showboxes]{textpos} 

\usepackage{hyperref}

\usepackage{cleveref}

\graphicspath{{figures/}}

\copyrightyear{2023}
\acmYear{2023}
\setcopyright{rightsretained}
\acmConference[IMC '23]{Proceedings of the 2023 ACM Internet Measurement 
	Conference}{October 24--26, 2023}{Montreal, QC, Canada}
\acmBooktitle{Proceedings of the 2023 ACM Internet Measurement Conference 
	(IMC '23), October 24--26, 2023, Montreal, QC, Canada}
\acmDOI{10.1145/3618257.3624840}
\acmISBN{979-8-4007-0382-9/23/10}

\makeatletter
\gdef\@copyrightpermission{
	\begin{minipage}{0.3\columnwidth}
		\href{https://creativecommons.org/licenses/by/4.0/}{\includegraphics[width=0.90\textwidth]{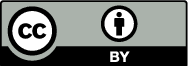}}
	\end{minipage}\hfill
	\begin{minipage}{0.7\columnwidth}
		\href{https://creativecommons.org/licenses/by/4.0/}{This work is 
			licensed under a Creative Commons Attribution International 4.0 
			License.}
	\end{minipage}
	\vspace{5pt}
}
\makeatother

\begin{document}

\title{Pushing Alias Resolution to the Limit}

\author{Taha Albakour}
\affiliation{\institution{TU Berlin}
    \country{}
}
\author{Oliver Gasser}
\affiliation{\institution{Max Planck Institute for Informatics}
    \country{}
}
\author{Georgios Smaragdakis}
\affiliation{
    \institution{Delft University of Technology}
    \country{}
}

\renewcommand{\shortauthors}{Taha Albakour, Oliver Gasser, \& Georgios 
	Smaragdakis}

\settopmatter{printacmref=true, printccs=true, printfolios=true}
\pagenumbering{gobble}

\begin{abstract}

In this paper, we show that utilizing multiple protocols offers a unique
opportunity to improve IP alias resolution and dual-stack inference
substantially. Our key observation is that prevalent protocols, e.g., SSH and
BGP, reply to unsolicited requests with a set of values that can be combined to
form a unique device identifier. More importantly, this is possible by just
completing the TCP handshake. Our empirical study shows that utilizing readily
available scans and our active measurements can double the discovered IPv4 alias
sets and more than 30$\times$ the dual-stack sets compared to the
state-of-the-art techniques. We provide insights into our method's accuracy and
performance compared to popular techniques.

\end{abstract}



\maketitle

\section{Introduction}\label{sec:intro}

Uncovering the Internet's topology is crucial for Internet measurement and analysis.
Common topology mapping tools, such as Traceroute, only provide partial
information by revealing interface-level links. {\it Alias resolution}, the process of
mapping IP addresses to the underlying hardware, enhances the accuracy and
completeness of the observed topology
~\cite{alias_importance}. Moreover, it can aid researchers in the
development of novel measurement techniques~\cite{reverse_tr}. 
The identification of {\it dual-stack} hosts, i.e, IPv4 and IPv6 enabled host,
presents a conceptually similar challenge to alias resolution. Due to its 
large address space, however, measuring IPv6 networks remains a challenging 
task. Nevertheless, identifying dual-stack hosts is an
important step in understanding network performance \cite{server-to-server},
policy \cite{ipv6-evo}, and security posture \cite{Back-Door-IPv6}. 

Prior work introduced many techniques to resolve aliases with the  
common source address~\cite{iffinder} as the earliest approach. This 
technique 
operates by sending a packet to a closed
port on a router, which triggers an ICMP port unreachable message. If the source
address of the ICMP message differs from the probed address  (the interface
where the packet is received), the IP pairs are inferred as aliases. However,
detecting aliases using this method becomes challenging as many routers always
respond from the probed address or may not respond at all, rendering the 
technique impractical.  

Other techniques utilize the IPID field in the IP header. 
IPID-based techniques are predicated on the fact that many routers maintain a
monotonic IPID counter that increments with each generated packet, and shared 
across interfaces. IPID-based tools attempt to sample the IPID value of 
candidate IPs over a short timeframe and perform a monotonic bounds test on 
the IPID sequences. If an IP pair share the same sequence, then they are 
likely to be aliases. RadarGun~\cite{RadarGun}, Rocketfuel~\cite{Rocketfuel},
and MIDAR~\cite{keys13midar} are 
few examples of tools utilizing this technique for IPv4 addresses, and 
Speedtrap~\cite{Speedtrap} for IPv6 addresses, respectively. If a router utilizes a non-monotonically 
incremental IPID counter, such technique fails to identify potential aliases. 
Additionally, these techniques require sending large number of packets, 
rendering them less optimal for large scale measurements.

Recent work took a protocol-centric approach and exploited a unique identifier
in the response to an unsolicited SNMPv3 request~\cite{IMC2021-SNMPv3}. This approach can infer aliases by
grouping addresses that shares the same unique identifier. One drawback to this
approach is that it requires the target IP to respond to a specific service, \ie SNMPv3.
Firewalls and access control lists can limit the number of identifiable aliases for a
given host if the service is configured to respond only on selected addresses. 

The mentioned techniques mainly addresses the alias resolution problem, 
however, the protocol-centric approach also solve the dual-stack 
identification.  
Further, researchers have developed a use-case specific solutions for dual-stack 
identification 
~\cite{IMC2021-SNMPv3,dual-stack-PAM2017,dual-stack-CoNEXT2022,IPv4-IPv6-Relationship:IMC2013}
with generic techniques utilizing DNS PTR records \cite{regex19, 
Back-Door-IPv6}

In this paper, we take a protocol-centric approach and introduce a technique 
that improves both IP alias
and dual-stack resolution.
Our main contributions can be summarized as follows:

\begin{itemize}[leftmargin=*]

\item We introduce a new alias resolution technique, for both IPv4 and IPv6, 
by
  collecting and analyzing application layer headers for different protocols,
namely, BGP and SSH.

\item Our alias resolution technique improves dual-stack discovery as more IPv4
  and IPv6 addresses are associated with unique identifiers.

\item We apply our methodology on our own active measurements data as well as 
data obtained from
    Censys. 
We complement previous protocol-centric technique and demonstrate that it 
is possible to more than double the number of identifiable non-singleton IPv4 
alias sets.

\item Our results show that we can identify more than 650 thousand dual-stack
alias sets. Which is, by a large margin, the largest set reported to date.

\item We make the datasets we collected and our analysis publicly available
at:~\url{https://routerfingerprinting.github.io/}

\end{itemize}

 \section{Methodology}\label{sec:methodology}

Scanning for active services is a widely used technique in Internet measurement
and security analysis~\cite{ZMap,Back-Door-IPv6}. In this paper, we show that
utilizing service scanning results for two popular protocols, namely, SSH and
BGP, enables large-scale alias and dual-stack inference. By analyzing these
protocols and their specifications~\cite{rfc4253, IETF-RFC4271}, we identify
unique host identifiers that can be used to group IP addresses belonging to the
same host in both IPv4 and IPv6.

\subsection{Service Scan Data}

We perform active service scans for SSH and BGP in two phases:

\begin{enumerate}[leftmargin=*]

\item An Internet-wide TCP scan sending a single SYN packet on port 22 and 179 
using ZMap \cite{ZMap}.

\item A service scan using ZGrab2~\cite{zgrab2} targeting IPs, 
which are responsive 
to the Internet-wide ZMap scan.

\end{enumerate}

In the service scan, specifically for SSH, we complete the TCP handshake and 
subsequently send a 
protocol-specific payload to solicit banner information from the target IP.
For BGP, the target IP sends an open message after we complete the TCP 
handshake without the need for any additional data exchange.
 
To complement our view of active services, we leverage the Censys dataset~\cite{Censys}, in addition to our own active measurements.
Censys perform service scan on the 65k ports. However, we only consider hosts 
that are running SSH and BGP on the default ports, i.e., TCP/22 for SSH and TCP/179 
for BGP.

\begin{figure}[t]
	\centering
	{\scriptsize
		\begin{Verbatim}[commandchars=\\\{\},frame=single]
SSH Protocol
\textcolor{blue}{Protocol: SSH-2.0-}
\textcolor{blue}{SSH Version 2}
 ...
 Key Exchange (method:curve25519-sha256)
  Message Code: Key Exchange Init (20)
  Algorithms
   ...
   \textcolor{blue}{kex_algorithms string: curve25519-sha256,..}
   server_host_key_algorithms length: 57
   \textcolor{blue}{server_host_key_algorithms string:...}
   encryption_algorithms_server_to_client length: 108
   \textcolor{blue}{encryption_algorithms_server_to_client string: ...}
   ...
   mac_algorithms_server_to_client length: 213
   \textcolor{blue}{mac_algorithms_server_to_client string: ...}
   ...
   compression_algorithms_server_to_client length: 21
   \textcolor{blue}{compression_algorithms_server_to_client string:...}
 Key Exchange (method:curve25519-sha256)
  Message Code: Elliptic Curve Diffie-Hellman Key Exchange Reply(31)
  KEX host key (type: ssh-ed25519)
   ...
   EdDSA public key length: 32
   \textcolor{blue}{EdDSA public key:409fa737033d6a79a1130aff96ee5ee2c39a9...}
  ...
		\end{Verbatim}
	}
	\caption{Snippet of a Dissected SSH Connection Setup}
	\label{fig:ssh-identifier}
\end{figure}

\subsection{SSH Identifier}

The Secure Shell (SSH) protocol, initially introduced in 
RFC 4253~\cite{rfc4253}, provides a mechanism to establish a secure 
network connection. We utilize ZGrab2's SSH module, which handle the SSH handshake, to perform 
our service scan. 
Upon completion of the TCP handshake, the server and 
the client send their respective service string banner and then proceed 
to exchange a series of plain text message before transitioning to an
encrypted session. During this exchange, both the server and client 
communicate their respective capabilities regarding encryption, 
authentication, and compression algorithms. This exchange enables both endpoints 
to convey to the other the algorithms they support. RFC 4253~\cite{rfc4253}
states that each supported algorithm MUST be listed in order of preference, from
most to least. This requirement results in a signature that can be used to
identify the client and the server implementation~\cite{hassh-honeypot,hassh}. 
We use this information, and the service banner as the first part of our SSH 
host identifier. 

SSH server requires a pair of host keys. These keys are typically generate 
during the service setup. The client and server exchange the public key 
components during the connection setup phase. We use the server public key as 
the second part of our SSH identifier. While the SSH public key itself is 
likely to be unique per host, our active scan shows that 0.4\% of 
non-singleton hosts communicate different algorithmic capabilities. 
Therefore, combining the key with the host's algorithmic capabilities can 
enhance the uniqueness of the SSH identifier. We highlight (in blue) the various parts 
of our SSH identifier in a snippet of SSH connection setup in 
\Cref{fig:ssh-identifier}.

\subsection{BGP Identifier}

\begin{figure}[t]
	\centering
	{\scriptsize
		\begin{Verbatim}[commandchars=\\\{\},frame=single]
Border Gateway Protocol - OPEN Message
	Marker: ffffffffffffffffffffffffffffffff
	\textcolor{blue}{Length:37}
	Type: OPEN Message (1)
	\textcolor{blue}{Version: 4}
	\textcolor{blue}{My AS: 23456}
	\textcolor{blue}{Hold Time: 90}
	\textcolor{blue}{BGP Identifier: 148.170.0.33}
	Optional Parameters \textcolor{blue}{Length: 8}
	Optional Parameter: Capability
	    Parameter Type: Capability (2)
		  Parameter Length: 2
		  \textcolor{blue}{Capability: Route refresh capability (Cisco)}
		  \textcolor{blue}{Type: Route refresh capability (Cisco) (128)}
		  Length: 0
	Optional Parameter: Capability
	    Parameter Type: Capability (2)
		  Parameter Length: 2
		  \textcolor{blue}{Capability: Route refresh capability}
		  \textcolor{blue}{Type: Route refresh capability (2)}
		  Length: 0
Border Gateway Protocol - NOTIFICATION Message
	Marker: ffffffffffffffffffffffffffffffff
	Length: 21
	Type: NOTIFICATION Message (3)
	Major error Code: Cease (6)
	Minor error Code (Cease): Connection Rejected (5)
		\end{Verbatim}
	}
	\caption{A Dissected BGP OPEN Message}
	\label{fig:bgp-identifier}
\end{figure}

The BGP protocol is used to facilitate the exchange of routing information between BGP-speaking routers.
To that end, BGP speakers establish and maintain a TCP session, typically 
over port 179. 
When scanning for host running BGP, we complete the TCP handshake and 
wait for data. We simply close the connection after 2 seconds timeout, or  
after receiving any data. 
We find that more than 5.8M BGP speakers close the connection immediately 
after completing the TCP handshake. However, 364k IPs close the connection 
after sending an OPEN and a Notification message stating that the connection 
is rejected.
\Cref{fig:bgp-identifier} shows an example of a dissected BGP OPEN message from 
our service 
scan. 

The OPEN message of a BGP speaker contains multiple fields that, when 
combined, can serve as a globally unique identifier. The first 
notable field is the BGP identifier. The BGP 
identifier is used as part of a loop and collision prevention mechanism and 
defined in RFC 4271 \cite{IETF-RFC4271} as 4-octet unsigned integer that 
uniquely identifies the 
BGP speakers within an Autonomous System (AS). Moreover, it should have the 
same value for every 
local interface. The OPEN message also contains the Autonomous System Number 
(ASN) of a BGP speaker's network. The ASN is a globally unique number that is 
associated with a single AS \cite{IETF-RFC1930}.
Some OPEN messages may contain optional parameters field that indicate the  
supported capabilities \cite{IETF-RFC5492}. 
The additional fields within the OPEN message such as Length, Version, and 
Hold Time are host-wide, and shared across all interfaces. 
Combining the values of those fields results in a unique identifier that we 
use to group alias and dual stack addresses.  
We highlight (in blue) the relevant parts 
of the identifier in a dissected BGP message in \Cref{fig:bgp-identifier}. 

\subsection{Alias and Dual-Stack Inference}

For every IP that is responsive to the BGP and SSH service scan, we extract
the respective identifier. We group IP addresses that shares the same 
identifier into SSH and BGP alias sets, respectively. We group IPv4 and IPv6 
addresses that share the same identifier into dual-stack sets.

\subsection{Datasets}
\label{sec:methodology:sub:datasets}

We leverage two different types of datasets. First, we use active measurement 
data in the IPv4 and IPv6 Internet.
In IPv4, we perform Internet-wide scans for the SSH and BGP 
protocols using ZMap \cite{ZMap} and ZGrab2 \cite{zgrab2}.
In IPv6, we use an IPv6 Hitlist 
\cite{gasser2018clusters,zirngibl2022rustyclusters} to identify potentially 
active addresses in the vast IPv6 address space.  The active measurement 
data was collected on April 18, 2023, utilizing a single vantage point 
located in a data center in Germany. 
Our dataset, including our analysis, are publicly available~\cite{IMC2023-alias-artifacts}.
Second, we use data obtained from Censys \cite{Censys} to 
identify additional responsive hosts to SSH or BGP. We selected a Censys 
snapshot that closely matches the date of our active measurement, March 28, 
2023.

In \Cref{table:scan-results} we show an overview of these two datasets as 
well as the union, where applicable, of both sources.
In IPv4, we find that both Censys as well as our active scans cover a similar number of ASes for both SSH and BGP.
Censys does, however, find around 6M more IPs for SSH and 35k more IPs for 
BGP. 
This might be linked with Censys performing distributed measurements, which 
reduces the likelihood of triggering rate-limiting or intrusion detection 
system filters \cite{origin_of_scanning}. Further, censys also 
finds an additional 5.6M IPs running SSH on 60,806 different ports. We do not 
consider non-standard ports from Censys since our active scan only covers 
port 22.
The union of both IPv4 data sources provides additional coverage compared 
to 
just a single source, both with respect to the number of covered IPs as well 
as ASes.
Therefore, unless explicitly stated otherwise, we use the union of both data 
sources in the remainder of the paper for our IPv4 analysis.

In IPv6, our active scans find more than 1M SSH IPs and 67k BGP IPs. In 
contrast, Censys reports only 944 SSH IPs and no IPs for BGP. Further, the 
SSH IPs are running the service on a non-standard port, namely 80 and 443. We 
believe that 
the variation attribute to the IPv6 hitlists used. Due to its limited 
coverage, we exclude Censys IPv6 data from our analysis. However, 
as of August 15, 2023,  
Censys IPv6 snapshot reports more than 415k IPv6 addresses running SSH on 
port 22. We expect this number to increase overtime  
as Censys scans for IPv6 more rigorously. 

In addition to SSH and BGP services, we conduct an SNMPv3 scan for both IPv4 
and IPv6. We utilizing an already established methodology 
\cite{IMC2021-SNMPv3} to identify alias and dual-stack sets. We then use the 
results for validation purposes and as a 
supplement to our results.  The SNMPv3 data also serve as baseline for 
comparison. We note that Censys data primarily reports SNMPv2 hosts and 
does not seem to include any information on SNMPv3. Consequently, we do not 
include it as an additional source.

\begin{table}[!bpt]
	\caption{Service Scanning Dataset Overview}
	\label{table:scan-results}  
	{\small
		\resizebox{\columnwidth}{!}{
			\begin{tabular}{lllllrr}
				\toprule
                & \multicolumn{2}{c}{Active measurements} & \multicolumn{2}{c}{Censys} & \multicolumn{2}{c}{Union} \\
                \cmidrule(lr){2-3}
                \cmidrule(lr){4-5}
                \cmidrule(lr){6-7}
				Protocol &  \# IPs & \# ASN  & \# IPs & \# ASN &  \# IPs & \# ASN \\
				\midrule                
				SSH  &  15.9M  & 46.1k  & 21.7M &  47.6k & 24.4M & 
				48.9k \\
				BGP  & 364k  & 6.5k & 391k  & 7k & 409k & 7.5k \\
				SNMPv3 & 20.8M & 50.2k  &  n.a & n.a & n.a & n.a \\
				\midrule
				Union & 36.7M & 59.6k  & 22.1M & 48.5k & 24.7M & 49.7k \\
				\midrule  
				SSH (IPv6) &  1.01M  & 10.8k &  n.a & n.a & n.a & n.a \\
				BGP (IPv6) &  67k  & 3.1k &  n.a & n.a & n.a & n.a \\
				SNMPv3 (IPv6) & 337k & 10.8k &  n.a & n.a & n.a & n.a \\
				\midrule
				Union & 1.3M & 14.4k &   &  &  & \\
				\bottomrule	

	\end{tabular}}}
\end{table}

\subsection{Validation}
We take a cross-protocol validation 
approach and compare sets derived from IP addresses responsive to different 
protocol pairs. 
We also utilize MIDAR \cite{keys13midar} as an additional 
source for validation. 
Specifically, we test a random sample of 61k alias sets using MIDAR and check 
whether the 
resulting sets perfectly match the ones we identify with SSH.
We ensure that each sample set contains at most ten IPv4 
addresses to ensure completing the MIDAR run in a close time 
frame to the SSH service scan. 
We provide a 
summary of our validation results in 
\Cref{table:validation-sets} where we report the test sample size, 
the 
number of sets that exactly match, and the number of sets with 
mismatching IPs.

In cross-protocol validation, we initially compare the alias sets 
obtained from SSH and BGP. Our active scan data contains a total of 7.8k 
responsive addresses, common to both protocols. We identify 1.34k alias sets 
using SSH and 1.35k alias sets using BGP. The validation between SSH and BGP 
protocols shows that 96\% of the SSH sets have a perfect 
match with the BGP sets.

Next, we examine the results of SSH and SNMPv3 pairs. Our active scan data 
contains a total of 63k responsive addresses to both protocols, resulting in 
13.6k alias sets using SSH and 14.5k alias sets using SNMPv3. The validation 
between SSH and SNMPv3 protocols shows a 97\% agreement.

Finally, we compare the BGP and SNMPv3 pairs with 37k responsive addresses to 
both protocols. We identify 1.84k alias sets using BGP and 1.9k alias sets 
using SNMPv3. The validation between BGP and SNMPv3 shows a 95\% agreement.

When comparing our results with MIDAR, we focus solely on 
SSH-based alias sets due to the time required to run MIDAR against 
all alias sets.
We find 
that only 13\% of the sampled sets can be verified with MIDAR. This 
low coverage can be attributed to two reasons: (a) the majority of these 
addresses do not utilize an incremental IPID counter, or (b) targets with 
large traffic 
volume resulting in a high velocity IPID counter. 
MIDAR is able to verify 8.5k alias sets with a 96\% agreement with our SSH results. 
The remaining 4\% alias sets are split into two or three alias sets by MIDAR, 
while SSH groups them into a single set. We suspect that the disagreement  
can be attributed to IP churn given that the MIDAR run took three weeks to 
complete. It is also possible that some of these sets share the same host 
key.  

In summary, the validation results confirm that our technique has at least a 
95\% agreement with state-of-the-art.

\begin{table}[!bpt]
	\caption{Alias Sets Validation}
	\label{table:validation-sets}  
	{\small
		\resizebox{\columnwidth}{!}{
			\begin{tabular}{lllrr}
				\toprule
				& Sample size & Agree & Disagree &  \\
				\midrule                
				SSH-BGP     & 1.34k  & 1.29k & 53 & \\
				SSH-SNMPv3  & 13.6k & 13.2k & 398&  \\
				BGP-SNMPv3  & 1.84k & 1.76k & 87 &  \\
				SSH-MIDAR   & 8.5k & 8.1k & 366 & \\
				\bottomrule   
	\end{tabular}}}
\end{table}

\subsection{Limitations}
Our methodology provides the largest sets of alias and dual-stack addresses 
to date. However, we do note a few limitations:
\begin{itemize}[leftmargin=*]
\item First, our methodology relies on application-level data. As such, it is 
only 
applicable to IPs responsive to SSH and BGP. Firewalls and access control may 
block or restrict access to the these services which can limit the alias 
inference.
\item Second, in the case of BGP, BGP speakers can have a non-unique BGP 
identifier due to mis-configuration which can lead 
to incorrect inferences. 
\item Third, our defined SSH identifier, might not be unique in all cases.
It is in fact possible 
for multiple host to share the same identifier, \eg SSH servers can be 
shipped with factory-default keys \cite{gasser2014deeper,Ps-Qs}. 
It is unlikely for two different hosts to generate the exact same host key, 
however, unless an administrator chose to use the same key pair across 
multiple hosts. 
\item Lastly, our validation is limited by the relatively small 
number of overlapping sets with other techniques, the responsiveness of a 
service on all IPs in a given set, and the possibility of IP churn. 
\end{itemize}

 \section{Ethical Considerations}\label{sec:ethics}

For our active experiments we do our best to minimize additional load or harm 
on the destination devices. BGP, SSH, and SNMPv3 load is very low (only a few
packets per destination). Moreover, we randomly distribute our measurements 
over the address space for our experiment, ensuring that at most one packet 
reaches a target IP each second.  Furthermore, we coordinate with local 
network administrators to ensure that our scanning efforts do not harm the 
local or upstream network.  For the active scanning we use best current
practices~\cite{ZMap,partridge2016ethical,dittrich2012menlo} to ensure that 
our prober IP address has a meaningful DNS PTR record. Additionally, we show
information about our measurements and opt-out possibilities on a website of 
our scanning servers. During our active experiments, we did not receive any
complaints or opt-out requests.  
 \section{Analysis}\label{sec:analysis}

In this section we present our results, consisting of alias resolution 
and dual-stack statistics as well as AS-level analyses.

\subsection{Alias Resolution}

To identify alias sets, we group IP addresses with identical unique 
identifiers for SSH and BGP. We also supplement our findings 
with SNMPv3 as described in \cite{IMC2021-SNMPv3}. In \Cref{table:alias-sets} 
we report the number of non-singleton alias sets and the contribution of each 
individual protocol, 
data source, and the union of all.
In IPv4, the SSH active scan results in 505k alias sets, which cover over 
3.2M unique IPv4 addresses. Similarly, the Censys dataset results in 699k 
alias sets, covering more than 4.6M IPv4 addresses. Censys data provide a 
notable increase of 70\% and 80\% in the number of 
IPv4 addresses and resulting alias sets compared to the active measurement 
alone.
 
With BGP, both Censys and the active scan produce similar results, with 12k 
alias sets covering 175k IPv4 addresses. In contrast, our SNMPv3 scan results 
in 557k alias sets covering 6.1M IPv4 addresses. By consolidating these 
findings, we can effectively cover more than 11.8M IPv4 addresses. 

\begin{figure}[!bpt]
	\centering
	\includegraphics[width=1\linewidth]{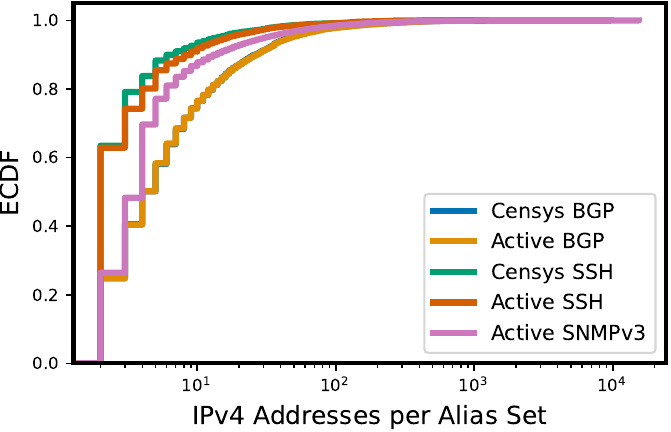}
\caption{IPv4 addresses per alias sets}
	\label{fig:alias-set-size}
\end{figure}

Interestingly, a substantial majority of 97\% of these addresses only respond 
to a single service, while only 3\% are responsive to two or three services. 
Consequently, this stark difference increases the 
resulting alias sets, exceeding 1.4M, of which 40\% can only be 
identified with SNMPv3 and 60\% (which is more than double what can be 
achieved by SNMPv3 alone) with SSH or BGP. We note however, that the majority 
of these sets comes from SSH. In 
\Cref{fig:alias-set-size} we show the 
distribution of IPv4 addresses per alias set. We find that the majority of 
the sets contain less than 100 addresses. Additionally, more that 60\% of SSH 
alias sets contain only two addresses compared to less than 30\% for BGP and 
SNMPv3. BGP sets are also more likely to contain more addresses compared to 
sets derived from SSH and SNMPv3. We also note a similar set size regardless 
of the data source.

\begin{figure}[!bpt]
	\centering
	\includegraphics[width=1\linewidth]{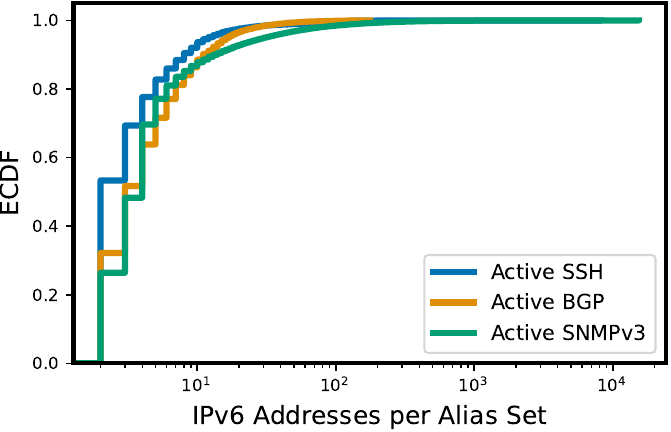}	
\caption{IPv6 addresses per alias sets}
	\label{fig:alias-set-size-ipv6}
\end{figure}

For IPv6, the active SSH scan results in 47k alias sets that cover 266k 
unique IPv6 addresses. Moreover, we find 8.3k and 16.7k alias sets, covering 
48k and 71k IPv6 addresses with BGP and SNMPv3, respectively. 
Merging these results
we obtain over 66k IPv6 alias sets, with a 
coverage of more than 340k unique IPv6 addresses. Similar to our IPv4 results, 
a majority of 94\% of these addresses are only responsive to a single 
service, while 6\% are responsive to two or three services. This results in 
25\% of the IPv6 alias sets being identifiable only with SNMPv3, while 75\% 
can be identified with SSH and BGP.
In \Cref{fig:alias-set-size-ipv6} we show the distribution of IPv6 addresses 
per 
alias set. Similar to IPv4, the majority of sets contain less that 100 
addresses. Additionally, SSH sets are more likely to contain fewer IPv6 
addresses compared to BGP and SNMPv3. We also note a similar set size for BGP 
and SNMPv3.

\begin{table}[!bpt]
	\caption{Alias Sets Overview}
	\label{table:alias-sets}  
	{\small
		\resizebox{\columnwidth}{!}{
			\begin{tabular}{llllrr}
				\toprule
                \multicolumn{2}{c}{Source} & Active (IPs) & Censys (IPs) &  
                Union (IPs)& 
                  \\
				\midrule                
            \ldelim\{{4}{*}[IPv4]                 & SSH        & 505k (3.2M) 
            & 699k (4.6M) &   926k (5.7M) &\\
                                                               & BGP        
                                                               & 12k (175k) 
                                                               & 12k (175k) 
                                                               &  12k 
                                                               (175k)&   \\
                                                               & SNMPv3     
                                                               & 557k (6.1M) 
                                                               & n.a &  557k 
                                                               (6.1M) 
                                                               &\\
                \cmidrule(lr){2-6}
                                                               & Union & 
                                                               1.04M & 708k 
                                                               & 1.4M 
                                                               (11.8M) &\\
				\midrule                
            \ldelim\{{4}{*}[IPv6]                                            
                   & SSH  & 47k (266k)  & n.a & n.a  &\\
                                                               & BGP  & 8.3k 
                                                               (48k) & n.a & 
                                                               n.a  &\\
                                                               & SNMPv3 & 
                                                               16.7k (71k) & 
                                                               n.a & n.a  &\\
                \cmidrule(lr){2-6}
                                                               & Union & 66k 
                                                               &  &  &\\
		\bottomrule		
	\end{tabular}}}
\end{table}

\subsection{Dual-Stack Inference}

Next, we shift our attention to the results of dual-stack identification, as 
summarized in \Cref{table:dual-stack}.
We merge alias sets from IPv4 and IPv6, if they use the same unique 
identifier.
The SSH active scan results in more 
than 634k dual-stack alias sets, which cover 
1.05M IPv4 addresses and 771k IPv6 addresses. With BGP, we identify 4.2k 
dual-stack sets, covering 78k IPv4 addresses and 16.3k IPv6 addresses. 
Additionally, SNMPv3  discovers 21k dual-stack alias sets that cover 1.1M IPv4 
addresses and 45k IPv6 addresses. Consolidating these findings results in a 
total of 650k dual-stack alias sets, of which 3\% can only be identified with 
SNMPv3, 
while 97\% (30$\times$ compared to SNMPv3 alone) can only be identified with 
SSH or BGP. Further, these sets cover 
a total of 2.2M IPv4 addresses and 830k IPv6 
addresses. 
Notably, more than 88\% of the dual-stack sets contains a single IPv4 and a 
single IPv6 addresses, 7\% set with 2-10 addresses, and only 2\% with 
more than 10 addresses. 
It is worth noting that our IPv6 sample size is relatively small
compared to IPv4. Nonetheless, these results indicate that a substantial
portion of known IPv6 addresses are exclusively IPv6-enabled, with just 64\% 
of the IPv6 addresses having an IPv4 counterpart. However, it is also 
possible that some host are only responsive over IPv6 due to policy as shown 
by previous work \cite{Back-Door-IPv6}.

\begin{table}[!bpt]
	\caption{Dual-Stack Sets}
	\label{table:dual-stack}  
	{\small
		\resizebox{\columnwidth}{!}{
			\begin{tabular}{llrr}
				\toprule
				           & IPv4 addr & IPv6 addr &  Dual Stack Sets \\
				\midrule                
				SSH        & 1.05M & 771k  & 634k \\
				BGP        & 78k & 16.3k & 4.2k \\				
				SNMPv3     & 1.1M & 45k & 21k  \\
				\midrule
				Union & 2.2M & 830k &  650k\\
				\bottomrule		
	\end{tabular}}}
\end{table}

\subsection{AS-Level Analysis}

\begin{figure}[!bpt]
	\centering
	\includegraphics[width=1\linewidth]{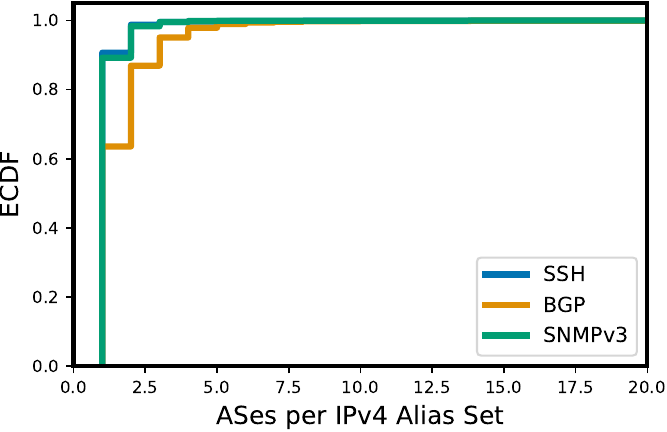}
\caption{ASN per IPv4 Alias Set}
	\label{fig:asn-per-alias-set}
\end{figure}

\Cref{fig:asn-per-alias-set} shows the distribution of Autonomous System 
Numbers (ASNs) per IPv4 alias set. We find that less than 10\% of SSH and 
SNMPv3 sets contain addresses associated with two or more ASes. In 
contrast, over 35\% of BGP sets contain addresses associated with multiple 
ASes. This outcome aligns with expectations, as BGP typically consist 
of border routers that connect different ASes.

\begin{figure}[!bpt]
	\centering
	\includegraphics[width=1\linewidth]{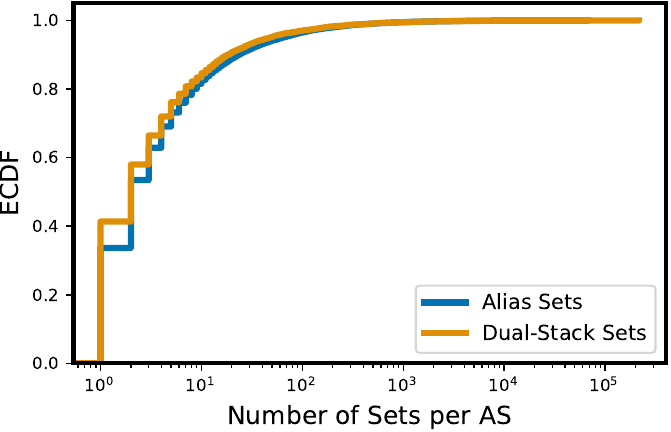}
\caption{Distribution of the number of alias sets per AS.}
	\label{fig:alias-set-per-as-set}
\end{figure}

In \Cref{fig:alias-set-per-as-set}, we show the distribution of the number 
of alias and dual-stack sets per AS. We find that over 37k ASes contain at 
least one set. The majority of ASes have fewer than 100 sets, and only 3\% of ASes 
have more than 100 alias sets.

\begin{table}[!bpt]
\caption{Top 10 ASes for IPv4 alias sets for each protocol separately and for the union. Each cell shows the ASN as well as the number of alias sets in parenthesis.}
\label{table:top-10-alias-sets}  
{\small
\resizebox{\columnwidth}{!}{
\begin{tabular}{lrrrr}
	\toprule
	Rank & SSH & BGP & SNMPv3 & Union \\
	\midrule
	1 & 14061 (68k) & 21859 (311) & 3269 (19k)  & 14061 (68k) \\
	2 & 22927 (61k) & 701 (288) & 30722 (12k) & 22927 (61k) \\
	3 & 16509 (46k) & 42689 (211) & 3320 (10k)  & 16509 (46k) \\
	4 & 16276 (28k) & 19429 (182) & 12874 (10k) & 16276 (29k) \\
	5 & 24940 (24k) & 24940 (162) & 4134 (8k) & 
	4134 (25k) \\
	6 & 14618 (23k) & 3269 (159) & 8881 (8k)  & 24940 (24k) \\
	7 & 45102 (19k) & 20473 (144) & 5089 (7.5k)& 3269 (23k) \\
	8 & 4134 (17k) & 12389 (144) & 3301 (7k)& 14618 (23k) \\
	9 & 396982 (17k) & 852 (101) & 7018 (7k) & 3320 
	(20k) \\
	10 & 46606 (15k) & 17511 (96) & 7029 (6.6k)  & 45102 (19k) \\
	\bottomrule
\end{tabular}}}
\end{table}

To better understand the main contributors of alias sets, we now focus on the top 10 ASes. In \Cref{table:top-10-alias-sets}, we 
report the largest AS based on different protocols as well as the 
union of all three protocols for IPv4.  
We expect SSH to be predominantly prevalent in cloud provider 
networks, whereas BGP and SNMPv3 to be more prevalent in ISP networks.
Indeed, among the top 10 ASes for SSH, 8 are cloud service providers, 
including DigitalOcean (rank 1, AS14061), Amazon (rank 3, AS16509; rank 6, AS14618), and OVH (rank 4, AS16276). 
Surprisingly, however, we also observe two major ISPs:
Telefonica de Argentina (rank 2, AS22927) and
China Telecom (rank 8, AS4134).
Shifting our focus to the top 10 ASes in the BGP and SNMPv3 data, we find
that 8 of them are ISPs, while the remaining 2 are cloud service providers.
The top three ASes for BGP are Zenlayer (AS21859), Verizon (AS701), and Glide (AS42689); the top three for SNMPv3 are Telecom Italia (AS3269), Vodafone Italy (AS30722), and Deutsche Telekom (AS3320).
Lastly, we consider the union of all data sources. We find this to be dominated by similar ASes as in the SSH data set, with a split of 6 
cloud service providers and 4 ISPs.

\begin{table}[!bpt]
    \centering
	\caption{Top 10 ASes for IPv6 alias and dual-stack sets. Each cell shows the ASN as well as the number of alias sets in parenthesis.}
	\label{table:top-10-alias-dual-sets}  
	{\begin{tabular}{lll}
                \toprule
				Rank & IPv6  & Dual-stack     \\
				\midrule
				1 &  7684 (4k) & 14061 (215k)\\
				2 &  63949 (3.2k) &  63949 (112k)\\
				3 &  4837 (1.8k) & 16276 (34k)\\
				4 &  4134 (1.4k) & 12876 (13k)\\
				5 &  6939 (1.1k) & 197695 (13k)\\
				6 &  26347 (943) & 8972 (8.7k)\\
				7 & 9808 (731) & 20473 (8.5k)\\
				8 &  197540 (713) &  8560 (8.5k)\\
				9 &  7922 (707)  & 7506  (7.7k)\\
				10 & 20857 (685) & 51167 (7.6k)\\
				\bottomrule
	\end{tabular}
}
\end{table}

We conclude our analysis by considering the largest 10 ASes with IPv6 
alias sets and dual-stack alias sets. 
\Cref{table:top-10-alias-dual-sets} shows the union results of all three 
protocols for IPv6 and IPv4-IPv6 dual-stack alias sets.
The IPv6 alias sets spread over 7k ASes in total.
The top 10 are split between 7 ISPs (\eg Hurricane Electric, AS6939; China 
Unicom, AS4837; Chinanet, AS4134)
and 3 cloud service providers (\eg Akamai, AS63949; Dreamhost, AS26347). 
Finally, our dual-stack alias sets cover more than 9.5k ASes. Note that this 
includes sets with at least a single IPv4 and a single IPv6 address. 
We find that the top 3 ASes are cloud service provides (DigitalOcean, ASAS14061; Linode, AS63949; OVH, AS16276)
and cover more than 54\% of the total dual-stack sets. The remaining 
7 are ISPs and cover only 10\% of all dual-stack alias sets.

 \section{Conclusion}\label{sec:conclusion}

In this paper we introduced a multi-protocol approach to improve IP alias resolution
and dual-stack identification. Our key observation is that a unique identifier
for each protocol can be used to group different subsets of alias sets. 
We evaluated
our method with two popular protocols, namely, SSH and BGP, and we showed
that our technique substantially increases both the number of alias as well as dual-stack sets,
compared to similar protocol-centric technique such as SNMPv3. Our results
showed that we can supplement previous work and identify up to 1.4 million 
non-singleton IPv4 alias
sets, i.e., double compared to what can be achieved with previously known
technique. Our results also showed that we can identify more than 650 thousand
dual-stack alias sets. By a large margin (30$\times$), this is the largest set
reported to date.

As part of our future research agenda, we plan to investigate if other popular
protocols are associated with unique identifiers that will further increase the
IP coverage of alias and dual-stack sets. We also plan to inspect
SSH identifiers more in-depth, specifically in terms of consistency and  
stability. Moreover, we plan to use updated IPv6 hit-list as we were
limited to these publicly available in this paper. Our initial results are very encouraging,
and we plan to perform additional measurements from multiple vantage points (VPs) to understand the effect of
geographical VP location.

 \section*{Acknowledgements}\label{sec:ack}

We would like to thank our shepherd, Liz Izhikevich, and the anonymous reviewers
for their valuable comments. This work was supported in part by the European
Research Council (ERC) under Starting Grant ResolutioNet (ERC-StG-679158).

\label{page:end_of_main_body}

\balance
\bibliographystyle{ACM-Reference-Format}

\label{page:last}

\end{document}